\documentclass[aps,prX,twocolumn,amsmath,amssymb,superscriptaddress,longbibliography,floatfix,raggedbottom]{revtex4-2}

\usepackage[colorlinks=true, linkcolor=blue, citecolor=blue, urlcolor=blue]{hyperref}
\usepackage[T1]{fontenc}
\usepackage[utf8]{inputenc}
\usepackage{bm}
\usepackage{natbib}
\usepackage{graphicx}
\usepackage{siunitx}
\usepackage{ulem}

\graphicspath{{figures/}{../figures/}}
\makeatletter
\newenvironment{widehere}[2][6pt]
  {%
    \def\wideheregap{#1}%
    \par\onecolumngrid
    \addvspace{\wideheregap}
    \noindent\begin{minipage}{\textwidth}
    \def\@captype{#2}%
    \small
    \centering
  }
  {%
    \end{minipage}\par
    \addvspace{\wideheregap}
    \twocolumngrid
  }
\makeatother
\begin{document}

\title{{Anatomy of orbital and spin responses in GaAs and Si: Hall, inverse Faraday, and magneto-optical Kerr effects}}
\author{Sanaz Alikhah}
 \email{sanaz.alikhah@physics.uu.se}
 \affiliation{Department of Physics and Astronomy, Uppsala University, P.O. Box 516, SE-75120 Uppsala, Sweden}

 \author{Daegeun Jo}
\affiliation{Department of Physics and Astronomy, Uppsala University, P.O. Box 516, SE-75120 Uppsala, Sweden}
\affiliation{Wallenberg Initiative Materials Science for Sustainability, Uppsala University, SE-75120 Uppsala, Sweden}

\author{Peter M. Oppeneer}
\affiliation{Department of Physics and Astronomy, Uppsala University, P.O. Box 516, SE-75120 Uppsala, Sweden}
\affiliation{Wallenberg Initiative Materials Science for Sustainability, Uppsala University, SE-75120 Uppsala, Sweden}

	\date{\today}
	
	\begin{abstract}
    
    Magneto-optical signals in nonmagnetic semiconductors entangle spin and orbital contributions, both in angular-momentum generation and magneto-optical detection. We calculate these contributions for electron-doped GaAs and Si using density-functional theory, linear-response theory, and a second-order density-matrix treatment of the inverse Faraday effect. At small electron doping, the intrinsic orbital Hall conductivity exceeds its spin counterpart in magnitude, while the orbital contribution to light-induced magnetization dominates over broad photon-energy ranges. We show that orbital- and spin-resolved optical Hall and polar Kerr coefficients exhibit stronger sensitivity to orbital moments than to equal spin moments. This contrast reflects direct electric-dipole coupling to orbital motion versus spin-orbit-mediated coupling to spin polarization. Consequently, Si maintains a substantial orbital magneto-optical response despite its weak spin-orbit interaction. At probe energies of 1.5 eV in GaAs and 3.0 eV in Si, the magnitude of the Kerr rotation per unit orbital moment is approximately 5 and 13 times larger than that per unit spin moment, respectively. Combining the calculated Hall conductivities and Kerr coefficients with a phenomenological drift-diffusion model yields predominantly orbital edge signals for the assumed diffusion constants and relaxation lengths. These results indicate that optical detection of induced magnetization in doped semiconductors such as GaAs and Si can be dominated by orbital signals, and therefore cannot be interpreted solely in terms of spin polarization.
    \end{abstract}

\maketitle
\section{Introduction}
\label{sec:introduction}

The ability to generate and transport orbital angular momentum (OAM) in solids has established orbitronics as a promising avenue beyond spin-based transport~\cite{go2021orbitronics, jo2024spintronics, cysne2025orbitronics, wang2025orbitronics, ando2025orbitronics, yesudas2026orbitronics,go2020orbital}. Its central appeal is that charge-to-orbital conversion can remain highly efficient even in materials with weak spin-orbit coupling (SOC)~\cite{go2018intrinsic, jo2018gigantic, salemi2022first,salemi2019orbitally,sala2022giant,tanaka2008intrinsic}, where the corresponding spin response is typically small. The intrinsic orbital Hall effect (OHE) was initially proposed for $p$-doped Si~\cite{bernevig2005orbitronics}. More recently, multiband calculations within the modern theory of orbital magnetization \cite{xiao2010berry} predicted large orbital Hall conductivities in spin-$3/2$ hole-doped semiconductors, including Si and GaAs~\cite{cullen2026orbital}, while a large room-temperature orbital Hall response has been reported in $n$-type Si~\cite{matsumoto2025observation}. These developments place conventional semiconductors among the material platforms now being explored for orbitronics.

This progress raises a corresponding question for magneto-optical detection. Kato \textit{et al.}\ used scanning Kerr microscopy to map an electrically induced edge signal in $n$-doped GaAs, which they attributed to spin accumulation generated by the spin Hall effect (SHE)~\cite{kato2004observation}. Kerr rotation, however, is not intrinsically spin selective and can respond even more strongly to an orbital moment than to a spin moment~\cite{choi2023observation,lyalin2023magneto}. Indeed, experiments in Ti~\cite{choi2023observation}, Cr~\cite{lyalin2023magneto}, Cu~\cite{ko2026magneto}, and Mn~\cite{ko2026electron} demonstrated that an orbital accumulation generated by the OHE dominated the observed Kerr signal. A spatial Kerr map therefore does not by itself determine the spin and orbital composition of the accumulated moment. Resolving this ambiguity requires determining both the spin-orbital composition of the edge accumulation and the Kerr response produced by each component.

The same spin-orbital distinction arises when angular momentum is generated optically \cite{adamantopoulos2025light,xu2021light}. Circularly polarized light can induce a bulk magnetization through the inverse Faraday effect (IFE), {a nonlinear opto-magnetic effect}~\cite{ziel1965optically, battiato2014quantum,pershan1966theoretical,hertel2006theory}. The first-principles theory of Berritta \textit{et al.}\ showed that the spin and orbital IFE contributions need not have the same magnitude,{ sign,} or photon-energy dependence~\cite{berritta2016ab}. Recent predictions of optically induced orbital polarization in bulk GaAs further motivate the study of optical orbital responses in semiconductors~\cite{scali2026optically}, although optical carrier injection and the coherent IFE considered here are distinct processes. The Hall effects and the IFE thus provide distinct means of generating angular momentum, one electrically and one optically, and comparing them allows us to examine whether orbital dominance is specific to one driving mechanism or is a broader property of the semiconductor response.

GaAs and Si provide complementary test cases. GaAs lacks inversion symmetry \cite{dresselhaus1955spin}, has stronger SOC, and has a direct band gap, whereas Si is centrosymmetric, has weaker SOC, and has an indirect band gap \cite{yao2005sign}. Their comparison allows us to examine whether sizable orbital responses persist when the spin channel is weak and how the different interband energy scales enter the optical response.

Here, we use first-principles calculations to systematically investigate the electrical generation, optical generation, and magneto-optical detection of orbital and spin responses in GaAs and Si. For electrical generation, we compare the intrinsic OHE and SHE of GaAs and Si. For optical generation, we calculate the orbital and spin components of the IFE. We then determine the optical Hall and polar magneto-optical Kerr effect (P-MOKE) responses separately per unit orbital or spin moment, thereby isolating the optical readout of the two components. For the Hall-driven response, the moment-resolved Kerr coefficients are combined with a drift-diffusion model of the edge accumulation to estimate the corresponding lateral Kerr profiles.

The key result is a twofold orbital dominance. First, the orbital response is generally larger than the spin response for both electrical Hall generation and optical generation through the IFE. Second, an orbital moment produces a stronger optical Hall and Kerr response per unit moment over most of the investigated photon-energy range. The orbital channel is therefore favored both when the moment is generated and when it is detected. This behavior is especially clear in Si, where the spin response is weak while the orbital response remains sizable. These findings suggest that substantial orbital responses can be generated by both electrical and optical means in semiconductors and highlight the potential of magneto-optical measurements to probe their signatures, motivating further experimental investigations of orbital transport in these systems.

\section{Methods}
\label{sec:methods}

\subsection{Computational details}
\label{sec:electronic_structure}

The electronic structures of bulk zinc-blende GaAs and diamond Si were calculated within density-functional theory (DFT) \cite{hohenberg1964inhomogeneous,kohn1965self}, using the all-electron full-potential linearized augmented-plane-wave method implemented in WIEN2k~\cite{blaha2020wien2k}.  The Perdew–Burke–Ernzerhof (PBE) generalized gradient approximation was used for exchange and
correlation~\cite{perdew1996generalized}, and spin–orbit coupling (SOC) was included in all calculations.
We used experimental cubic lattice constants of
$5.653~\text{\AA}$ for GaAs~\cite{blakemore1982semiconducting}
and $5.431~\text{\AA}$ for Si~\cite{okada1984silicon}.
Electron doping was simulated by adding $\Delta N_e=4\times10^{-5}$ electrons per two-atom primitive cell, together with a compensating homogeneous background charge. 
%Using $V_{\rm prim}=a^3/4$, this
{This} corresponds to carrier densities of approximately $8.86\times10^{17}~\mathrm{cm}^{-3}$ for GaAs and $9.99\times10^{17}~\mathrm{cm}^{-3}$ for Si. The product of the smallest muffin-tin radius $R_{\mathrm{MT}}$ and the largest reciprocal vector $K_{\mathrm{max}}$ for these systems was $R_{\mathrm{MT}} \times K_{\mathrm{max}} = 8.5$. A $27\times27\times27$ Monkhorst–Pack \cite{aspnes1983dielectric} grid was used for the self-consistent calculation, which converged to the nonmagnetic reference state, and a $58\times58\times58$ grid was used for the linear-response calculations. The conventional cubic cell was employed with crystal axes aligned with the Cartesian axes.
%with $x\parallel[100]$, $y\parallel[010]$, and $z\parallel[001]$.

\subsection{Linear-response calculations}

Using the calculated DFT electronic structures, we evaluated the {Hall and} magneto-optical 
%spectra 
{responses} within the linear-response formalism~\cite{kubo1957statistical,oppeneer2001,salemi2021quantitative,oppeneer1992ab}. Under the influence of an external perturbation $\hat{V} = -e\,\hat{\bm r}\cdot\bm{E}$, where $e$ is the electron charge, $\bm{E}$ is an external electric field of angular frequency $\omega$, and $\hat{\bm r}$ is the position operator, the response of a vector observable represented by the operator $\hat{\bm A} = (\hat{A}_x, \hat{A}_y, \hat{A}_z)$ can be expressed as $\delta A_i = \sum_{j} \chi^A_{ij}\,E_j $.
The response function $\chi^A_{ij}$ is expressed in terms of solutions of the Kohn-Sham Hamiltonian:
\begin{eqnarray}
		\chi^A_{ij}(\omega)\!\!&=\!\!& -\frac{i e}{m_e} \!\int_{BZ} \! \frac{d^3{k}}{\Omega}
		\sum_{n\neq m} \! \frac{f_{n\bm{k}} - f_{m\bm{k}}}{\hbar\omega_{nm\bm{k}}}
		\frac{A^i_{mn\bm{k}}\,p^j_{nm\bm{k}}}{\omega - \omega_{nm\bm{k}} + i\,\tau^{-1}_{\mathrm{inter}}} \nonumber
        \\
		&&- \frac{i e}{m_e} \!\int_{BZ} \!\frac{d^3\bm{k}}{\Omega}
		\sum_{n} \! \frac{\partial f_{n\bm{k}}}{\partial\mathcal{E}}
		\frac{A^i_{nn\bm{k}}\,p^j_{nn\bm{k}}}{\omega + i\,\tau^{-1}_{\mathrm{intra}}}.
	\label{eq:response}
\end{eqnarray}  
Here, $m_e$ is the electron mass, $f_{n\bm{k}}$ is the occupation of Kohn–Sham state $\lvert n\bm{k}\rangle$, $\Omega$ is the Brillouin-zone volume, $p^j_{nm\bm{k}}$ is the momentum-operator matrix element, $A^i_{mn\bm{k}}$ is the $\hat{\bm{A}}$-operator matrix element, and $\hbar\omega_{nm\bm{k}}=\mathcal{E}_{n\bm{k}}-\mathcal{E}_{m\bm{k}}$ is the difference of eigenenergies. The first term in Eq.\ (\ref{eq:response}) gives the interband (or Fermi sea) contribution, the second term the intraband (or Fermi surface) contribution. The quantities $\tau_{\mathrm{inter}}$ and $\tau_{\mathrm{intra}}$ are the electronic lifetimes for inter{band} and intraband transitions, respectively. 
We define the interband and intraband energy broadenings as $\Gamma_{\mathrm{inter}}=\hbar{\tau^{-1}_{\mathrm{inter}}}$ and $\Gamma_{\mathrm{intra}} ={\hbar}{\tau^{-1}_{\mathrm{intra}}}$.
In this work, we set $\Gamma_{\mathrm{inter}}=\Gamma_{\mathrm{intra}}=0.1~\mathrm{eV}${, unless otherwise specified}.

\subsection{Intrinsic spin and orbital Hall conductivities}
\label{sec:hall_method}

Following Ref.~\cite{salemi2022first}, the conventional spin- and orbital-current-density operators are {defined as}
\begin{equation}
    \hat{J}_i^{S_k}
    =\frac{\{\hat S_k,\hat p_i\}}{2V\,m_e},
    \qquad
    \hat{J}_i^{L_k}
    =\frac{\{\hat L_k,\hat p_i\}}{2V\,m_e},
    \label{eq:hall_operators}
\end{equation}
where $\{\hat A,\hat B\}=\hat A\hat B+\hat B\hat A$ denotes the anticommutator, $\hat{S}_k$ ($\hat{L}_k$) is the $k$th component of the spin (orbital) angular momentum operator, $\hat{p}_i$ is the $i$th component of the momentum operator, and $V$ is the volume of the unit cell. Substituting $\hat A_i=\hat{J}_i^{X_k}$, with $X=S$ or $L$, in Eq.~\eqref{eq:response} and taking $\omega=0$ gives {the dc spin or orbital conductivity}
\begin{equation}
    \sigma_{ij}^{X_k}
    =-\frac{ie}{m_e}\int_{\rm BZ}\frac{d^3k}{\Omega}
    \sum_{n\ne m}
    \frac{f_{n\bm k}-f_{m\bm k}}{\hbar\omega_{nm\bm k}}
    \frac{J_{i,mn\bm k}^{X_k}p^j_{nm\bm k}}
    {-\omega_{nm\bm k}+i\tau_{\rm inter}^{-1}}.
    \label{eq:hall_kubo}
\end{equation}
%Equation~\eqref{eq:hall_kubo} is the dc transverse-conductivity formula. For the transverse components considered here, the Brillouin-zone integral of the intraband term in Eq.~\eqref{eq:response} vanishes by time-reversal and cubic symmetry; consequently, only the interband term contributes. 
{The intraband term in Eq.~\eqref{eq:response} vanishes due to time-reversal symmetry.}
For a nonmagnetic cubic crystal, {symmetry allows only the Hall components} $\sigma_{ij}^{S_k}=\epsilon_{ijk}\sigma_{\mathrm{SH}}$ and
$\sigma_{ij}^{L_k}=\epsilon_{ijk}\sigma_{\mathrm{OH}}$ \cite{seemann2015symmetry}.
%The components used here are
{Consequently, we evaluate the spin and orbital Hall conductivities}
$\sigma_{\mathrm{SH}}=\sigma_{xy}^{S_z}$ and
$\sigma_{\mathrm{OH}}=\sigma_{xy}^{L_z}$, so that
$J_x^{S_z}=\sigma_{\mathrm{SH}}E_y$ and
$J_x^{L_z}=\sigma_{\mathrm{OH}}E_y$.

\subsection{Inverse Faraday effect}
\label{sec:ife_method}

We calculate the inverse Faraday effect using the second-order density-matrix formalism employed in Ref.~\cite{berritta2016ab}. In this formalism, the second-order density matrix $\hat{\rho}^{[2]}$ is obtained from the Liouville-von Neumann equation, and the magnetization {is} induced along the propagation direction of the light, {i.e., $\bm{M}_{ind} \propto i \bm{E} \times \bm{E}^{\star}$.} 
{Its magnitude is given by}
$M_{\mathrm{ind}} = -\mu_{\mathrm B}\operatorname{Tr} \{(\hat L_z+2\hat S_z)\hat{\rho}^{[2]}\}$, where $\mu_\mathrm{B}$ is the Bohr magneton.

For circularly polarized light of frequency $\omega$ propagating along the $z$ direction, the induced magnetization per unit volume is
\begin{equation}
    M_{\mathrm{ind}}
    =
    \left(
    K_{\mathrm o}
    +K_{\mathrm{dA}}
    +K_{\mathrm{dB}}
    +\mathrm{c.c.}
    \right)E_0^2,
    \label{eq:ife_second_order}
\end{equation}
where $E_0$ is the electric-field amplitude. 
%The magnetic moment is induced along the wave vector of the light. 
The three contributions are given by
% \begin{equation}
% \begin{split}
% K_{\mathrm o}
% ={}&
% \frac{e^2}{m_e^2\omega^2}
% \!\!\sum_{n\neq m;l}\!\!
% M_{mn}
% \frac{
% \dfrac{
% p^+_{nl}p^-_{lm}(f_m-f_l)
% }{
% \mathcal E_l-\mathcal E_m+i\Gamma_{lm}-\hbar\omega
% }
% -
% \dfrac{
% p^-_{nl}p^+_{lm}(f_l-f_n)
% }{
% \mathcal E_n-\mathcal E_l+i\Gamma_{nl}-\hbar\omega
% }
% }{
% \mathcal E_n-\mathcal E_m+i\Gamma_{nm}
% },
% \nonumber
% \end{split}
% \end{equation}
\begin{align}
K_{\mathrm o}
={}&
\frac{e^2}{m_e^2\omega^2}
\sum_{n\neq m;l}
\frac{M_{mn}}
{\mathcal E_n-\mathcal E_m+i\Gamma_{nm}} \nonumber
\\
&\times\!
\Bigg[
\frac{
p^+_{nl}p^-_{lm}(f_m-f_l)
}{
\mathcal E_l-\mathcal E_m+i\Gamma_{lm}-\hbar\omega
} -
\frac{
p^-_{nl}p^+_{lm}(f_l-f_n)
}{
\mathcal E_n-\mathcal E_l+i\Gamma_{nl}-\hbar\omega
}
\Bigg], \nonumber \\
K_{\mathrm{dA}}
={}&
\frac{e^2}{m_e^2\omega^2}
\sum_{nl}M_{nn}
\Bigg[
\frac{
p^+_{nl}p^-_{ln}(f_l-f_n)
}{
\bigl(
\mathcal E_l-\mathcal E_n+i\Gamma_{ln}-\hbar\omega
\bigr)^2
} \nonumber
\\
&\qquad\qquad
+
\frac{
p^-_{nl}p^+_{ln}(f_l-f_n)
}{
\bigl(
\mathcal E_n-\mathcal E_l+i\Gamma_{nl}-\hbar\omega
\bigr)^2
}
\Bigg],
\nonumber \\
K_{\mathrm{dB}}
={}&
\frac{e^2}{m_e^2\omega^2}
\sum_{nl}
\frac{M_{nn}}{\hbar\omega}
\frac{
p^+_{nl}p^+_{ln}(f_n-f_l)
\bigl(i\Gamma_{ln}-\hbar\omega\bigr)
}{
\bigl(\mathcal E_l-\mathcal E_n\bigr)^2
+
\bigl(\Gamma_{ln}+i\hbar\omega\bigr)^2
}.
\label{eq:ife_terms}
\end{align}
For brevity, the $\bm{k}$ dependence of all band-dependent quantities is suppressed (e.g., $\mathcal E_n\equiv\mathcal E_{n\bm k}$ and $f_n\equiv f_{n\bm k}$), and integration over the Brillouin zone is implied. Here, $p^\pm_{nl}=p^x_{nl}\pm i p^y_{nl}$ are circular components of the momentum-operator matrix elements, and
\begin{equation}
    M_{mn}
    =
    -\mu_{\mathrm B}
    \left(L^z_{mn}+2S^z_{mn}\right)
    \label{eq:ife_magnetization_matrix}
\end{equation}
is the magnetization matrix element (the wave vector $\bm k$ is implicit in the band indices and sums). The parameter $\Gamma_{nm}$ denotes the energy broadening entering the response denominators.

The contribution $K_{\mathrm o}$ contains off-diagonal magnetization matrix elements, $M_{mn}$ with $n\neq m$, and describes electronic Raman scattering~\cite{battiato2014quantum,popova2012theoretical}. The terms $K_{\mathrm{dA}}$ and $K_{\mathrm{dB}}$ contain diagonal matrix elements $M_{nn}$ and describe Rayleigh scattering. Equation~\eqref{eq:ife_terms} corresponds to the response functions for right-circularly polarized light, whose helicity is denoted as $\sigma^+$. The results for left-circularly polarized light, with helicity $\sigma^-$, are obtained by interchanging $p^+\leftrightarrow p^-$ in Eq.~\eqref{eq:ife_terms}.

Because the magnetization operator contains both orbital and spin contributions, the induced magnetization can be separated as
\begin{equation}
    M_{\mathrm{ind}}
    =
    -\mu_{\mathrm B}
    \operatorname{Tr}
    \left(\hat L_z\hat{\rho}^{[2]}\right)
    -2\mu_{\mathrm B}
    \operatorname{Tr}
    \left(\hat S_z\hat{\rho}^{[2]}\right).
    \label{eq:ife_spin_orbital}
\end{equation}
The orbital and spin contributions are obtained from Eq.~\eqref{eq:ife_terms} by making the substitutions $M_{mn}\rightarrow-\mu_{\mathrm B}L^z_{mn}$ and $M_{mn}\rightarrow-2\mu_{\mathrm B}S^z_{mn}$, respectively, with the same substitutions applied to the diagonal elements $M_{nn}$. Writing
$K(\omega)=K_{\mathrm o}+K_{\mathrm{dA}}+K_{\mathrm{dB}}$, the induced magnetization is expressed as
\begin{equation}
\begin{split}
M_{\mathrm{ind}}
&=
2\operatorname{Re}\left[K(\omega)E_0^2\right]
\equiv K_{\mathrm{IFE}}(\omega)\frac{I}{c}
\\
&=
\left[
K_{\mathrm{IFE}}^\ell(\omega)
+K_{\mathrm{IFE}}^s(\omega)
\right]\frac{I}{c}.
\end{split}
\label{eq:ife_definition}
\end{equation}
Here, $K_{\mathrm{IFE}}^\ell(\omega)$ and
$K_{\mathrm{IFE}}^s(\omega)$ are the orbital and spin
contributions to the IFE coefficient, respectively.
The light intensity and the total IFE coefficient are defined as
\begin{equation}
    I=\frac{\varepsilon_0cE_0^2}{2},
    \qquad
    K_{\mathrm{IFE}}(\omega)
    =
    \frac{4}{\varepsilon_0}
    \operatorname{Re}\left[K(\omega)\right].
    \label{eq:ife_coefficient}
\end{equation}
With this definition, $K_{\mathrm{IFE}}$ has units of T$^{-1}$. We used an energy broadening $\Gamma_{nm}=0.1$~eV and evaluated the Brillouin-zone sums on the $58\times58\times58$ $k$ mesh.

\subsection{{Spin- and orbital-resolved magneto-optical responses}}
\label{sec:moment_sources}

Spin and orbital moments generated in the system can be probed by magneto-optical responses \cite{Stamm2017magneto,choi2023observation,lyalin2023magneto,alikhah2026theory,marui2023spin}.
The frequency-dependent optical conductivity tensor $\bm{\sigma}(\omega)$ is obtained from Eq.~\eqref{eq:response} by replacing $\hat A_i$ with the $i$th component of the charge-current-density operator. In Gaussian units, the frequency-dependent dielectric tensor is
\begin{equation}
	\varepsilon_{ij}(\omega)
	=
	\delta_{ij}
	+
	\frac{4\pi\mathrm{i}}{\omega}\,
	\sigma_{ij}(\omega).
    \label{eq:dielectric}
\end{equation}
In the nonmagnetic system, the off-diagonal, antisymmetric optical conductivity vanishes by time-reversal symmetry. {However, we consider the nonequilibrium state in which the time-reversal symmetry is broken due to the spin or orbital accumulation induced by external fields, for example, by the spin or orbital Hall effect.} Following the procedure used in Ref.~\cite{lyalin2023magneto}, we model 
%small
{weakly} induced spin and orbital moments by self-consistently including the Zeeman term
\begin{equation}
	\hat{H}_\mathrm{ind}
	=
	\mu_{\rm B}B
	\left(\alpha\hat{L}_z+2\beta\hat{S}_z\right),
	\label{eq:zeeman_probe}
\end{equation}
in the 
%relativistic 
{nonmagnetic} Kohn-Sham Hamiltonian. Here,
%, $\hat{S}_z$ and $\hat{L}_z$ are the spin and orbital angular-momentum operators along $z$, and $B$ is the numerical probe-field amplitude. The dimensionless parameters $\alpha$ and $\beta$ control the coupling to the orbital and spin angular momenta, respectively. 
{$B$ is the effective magnetic field that couples to spin and orbital angular momenta with coupling strengths controlled by the dimensionless parameters $\beta$ and $\alpha$, respectively.} We use $(\alpha,\beta)=(0,1)$ for a spin-only probe and $(\alpha,\beta)=(1,1)$ for a spin-plus-orbital probe.
We label these two probe settings by $p=1$ and $p=2$, respectively.

For each probe setting $p=1,2$, the odd-in-field part of the off-diagonal optical conductivity is obtained by subtracting its value at a reversed probe field, i.e.,
\begin{equation}
	\Delta\sigma^{(p)}_{ij}(\omega;B)
	=
	\frac{
		\sigma^{(p)}_{ij}(\omega;B)
		-
		\sigma^{(p)}_{ij}(\omega;-B)
	}{2}.
	\label{eq:odd_in_B_probe}
\end{equation}
This antisymmetrization isolates the leading-order response linearly proportional to $B$ while eliminating even-in-B contributions. The induced moments are obtained in the same way,
\begin{equation}
	\delta m_q^{(p)}(B)
	=
	\frac{
		m_q^{(p)}(B)
		-
		m_q^{(p)}(-B)
	}{2},
	\qquad q=s,\ell.
	\label{eq:odd_moments_alpha}
\end{equation}
Here, $\delta m_s^{(p)}$ and $\delta m_\ell^{(p)}$ represent the linear contributions to the induced spin and orbital magnetic moments $m_s^{(p)}$ and $m_\ell^{(p)}$, respectively, for probe setting $p$, in units of $\mu_{\rm B}$.
Because SOC mixes spin and orbital character, the spin-only probe can also produce a small orbital moment through SOC, while the spin-plus-orbital probe induces both moments directly via Zeeman coupling.

In the small-moment regime, the odd-in-field optical Hall conductivity is linear in the induced spin and orbital moments. In the coordinate convention used here, with the induced moment along the $z$ axis, the relevant component is $\sigma_{xy}(\omega)$, and for each probe setting we write

 \begin{equation}
    \Delta\sigma^{(p)}_{xy}(\omega;B)
     \simeq
    \sigma^s_{xy}(\omega)\,\delta m_s^{(p)}(B)
     +
    \sigma^\ell_{xy}(\omega)\,\delta m_\ell^{(p)}(B),
     \label{eq:moment_linearization}
 \end{equation}
where $\sigma^s_{xy}(\omega)$ and $\sigma^\ell_{xy}(\omega)$ describe the optical Hall response per induced spin and orbital moment, respectively. 
Writing Eq.~\eqref{eq:moment_linearization} for the two probe settings gives
\begin{equation}
\begin{pmatrix}
    \Delta\sigma_{xy}^{(1)}(\omega;B)\\
    \Delta\sigma_{xy}^{(2)}(\omega;B)
\end{pmatrix}
=
\begin{pmatrix}
    \delta m_s^{(1)}(B) & \delta m_\ell^{(1)}(B)\\
    \delta m_s^{(2)}(B) & \delta m_\ell^{(2)}(B)
\end{pmatrix}
\begin{pmatrix}
    \sigma_{xy}^{s}(\omega)\\
    \sigma_{xy}^{\ell}(\omega)
\end{pmatrix}.
\label{eq:moment_inversion}
\end{equation}
The coefficients $\sigma_{xy}^{s}(\omega)$ and $\sigma_{xy}^{\ell}(\omega)$ are obtained by solving this $2\times2$ system independently at each photon energy. We used $|B|=10~\mathrm{T}$ and verified that both $\delta m_q$ and $\Delta\sigma_{xy}$ are linear in $B$ over the range $10-100~\mathrm{T}$. When the optical conductivity is reported in Gaussian frequency units, $\sigma^s_{xy}$ and $\sigma^\ell_{xy}$ have units of $\mathrm{s^{-1}\,\mu_{\rm B}^{-1}}$.

\subsection{{Magneto-optical Kerr effects due to induced spin and orbital moments}}
\label{sec:kerr_method}

The calculated optical conductivities are used to evaluate the spin- and orbital-resolved MOKE. For normal incidence polar MOKE, where $z$ is the surface normal, the complex Kerr coefficients per induced spin and orbital magnetic moments are~\cite{oppeneer2001}
\begin{align}
	\Phi_{K}^{q}(\omega)
	&\equiv
	\theta_{K}^{q}(\omega)
	+
	i \varepsilon_{K}^{q}(\omega)
	\nonumber\\
	&\simeq
	-\frac{
		\sigma^q_{xy}(\omega)
	}{
		\sigma_{xx}(\omega)
		\left[
		1+\dfrac{4\pi i}{\omega}\sigma_{xx}(\omega)
		\right]^{1/2}
	},
	\qquad q=s,\ell.
	\label{eq:pmoke}
\end{align}
which have units of rad $\mu_\mathrm{B}^{-1}$. The diagonal and off-diagonal optical conductivities entering Eq.~\eqref{eq:pmoke} were evaluated following the methods described in Sec.~\ref{sec:moment_sources}.

\section{Results}
\label{sec:results}

\subsection{Diagonal dielectric response and electronic structure}

We first examine the diagonal dielectric function $\varepsilon_{xx}(\omega)$ in the nonmagnetic ground states of electron-doped GaAs and Si. Figures~\ref{fig:epsilon}(a) and \ref{fig:epsilon}(b) show the real and imaginary parts of $\varepsilon_{xx}$ for GaAs and Si, respectively. Here, $\varepsilon_1$ and $\varepsilon_2$ denote the real and imaginary parts, respectively. The corresponding band structures are shown in Figs.~\ref{fig:epsilon}(c) and \ref{fig:epsilon}(d). The calculation includes both the interband and intraband terms of Eq.~\eqref{eq:response}. Electron doping enters through the free-carrier intraband contribution and through the occupation factors near the conduction-band minima. The pronounced structures at higher photon energies are dominated by interband transitions and reflect the joint density of states weighted by the optical matrix elements. The corresponding dispersive structures in $\varepsilon_{1}$ follow from the Kramers-Kronig relation \cite{ambrosch2006linear}.

In GaAs, the direct gap at $\Gamma$ determines where the interband response begins to rise above the low-energy background. The calculated $\varepsilon_{2}$ exhibits a broad structure near $2$--$2.5$~eV and a stronger resonance near $4$--$4.5$~eV. The strong high-energy absorption is accompanied by a rapid dispersive decrease and subsequent sign change of $\varepsilon_{1}$.

In Si, the valence-band maximum occurs at $\Gamma$, whereas the conduction-band minimum lies along $\Gamma$--$X$, making the fundamental gap indirect. Since phonon-assisted transitions, which govern the indirect absorption onset of Si, are not included~\cite{noffsinger2012phonon,zacharias2015stochastic}, the indirect gap does not produce a distinct feature in the calculated response. Instead, the increase of $\varepsilon_{2}$ near $2.5$~eV marks the onset of direct vertical transitions in the calculated spectrum.

The calculated spectra reproduce qualitatively the ordering and overall line shapes 
experimentally measured for both GaAs~\cite{aspnes1983dielectric,jellison1992optical} and Si~\cite{aspnes1983dielectric}.
This comparison concerns the main interband structures rather than the low-energy response. In particular, the Si sample studied in Ref.~\cite{aspnes1983dielectric} had an electron density of $2.3\times10^{14}~\mathrm{cm}^{-3}$, substantially lower than approximately $10^{18}~\mathrm{cm}^{-3}$ considered here. Moreover, experimental Si includes phonon-assisted indirect absorption, which is absent from the present calculation. Differences in the low-energy response are therefore expected.

The calculated spectral features are shifted toward lower photon energies relative to experiment, because the PBE exchange-correlation functional underestimates the relevant interband energy separations~\cite{borlido2020exchange}. Quasiparticle corrections primarily modify the transition energies \cite{hybertsen1986electron}, whereas local-field and electron-hole effects redistribute the optical spectral weight \cite{alouani1996optical,arnaud2001local,rohlfing2000electron,hanke1980many,Sagredo2025}. These effects are not included in the present calculations, which may contribute to the quantitative differences in peak positions and amplitudes.  Nonetheless, the qualitative agreement in the main spectral features supports the use of the optical input for the subsequent magneto-optical calculations in Sec. \ref{sec:spin_orbital_Hall_response}.

\begin{widehere}{figure}
    \centering
    \includegraphics[width=0.38\textwidth]{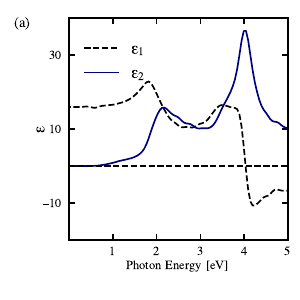}
\hspace{-2mm}
\includegraphics[width=0.38\textwidth]{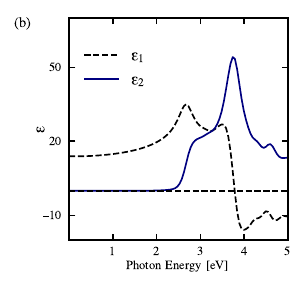}
\par\vspace{-5mm}
\includegraphics[width=0.38\textwidth]{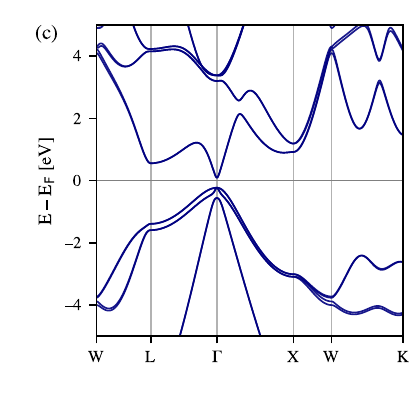}%
\hspace{-2mm}%
\includegraphics[width=0.38\textwidth]{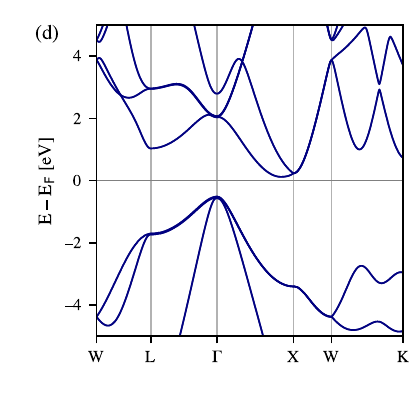}
\vspace*{-7mm}
    \caption{Calculated real ($\varepsilon_1$, dashed) and imaginary ($\varepsilon_2$, solid) parts of the diagonal dielectric function for electron-doped (a) GaAs and (b) Si, together with the corresponding PBE band structures including spin-orbit coupling for (c) GaAs and (d) Si. Band energies are measured relative to the Fermi level.}
    \label{fig:epsilon}
\end{widehere}

\subsection{Intrinsic orbital and spin Hall effects}
\label{sec:hall_results}

We now compute $\sigma_{\mathrm{SH}}(\mu)$ and $\sigma_{\mathrm{OH}}(\mu)$ for electron-doped GaAs and Si, where $\mu$ denotes the chemical potential measured relative to the self-consistent Fermi energy (see Fig.~\ref{fig:hall}). At $\mu=0$, the orbital Hall conductivity is approximately $-191.2$ and $72.8$ for GaAs and Si, respectively, in units of $(\hbar/e)(\Omega\,\mathrm{cm})^{-1}$. The sizable plateau of the OHC and the negligible SHC within the band gap are consistent with the previous calculations for Si~\cite{baek2021negative, matsumoto2025observation} and transition metal dichalcogenides~\cite{bhowal2020intrinsic,canonico2020orbital}. It should be noted, however, that a finite OHC does not directly imply boundary orbital accumulation in an insulating phase. While the orbital current originates from the Fermi-sea response, the orbital accumulation originates from the dissipative Fermi-surface response that requires finite carriers, which can be induced by doping \cite{sun2025nonconserved}. In the weakly doped regime, the OHC still dominates over the SHC, suggesting sizable current-induced orbital accumulation in doped GaAs and Si. We further discuss the MOKE signal expected from the intrinsic OHE and SHE and show that the OHE provides the dominant contribution for both GaAs and Si (Sec.\ \ref{sec:Hall_induced_Kerr}).

\begin{widehere}{figure}
    \centering
\includegraphics[width=0.38\textwidth]{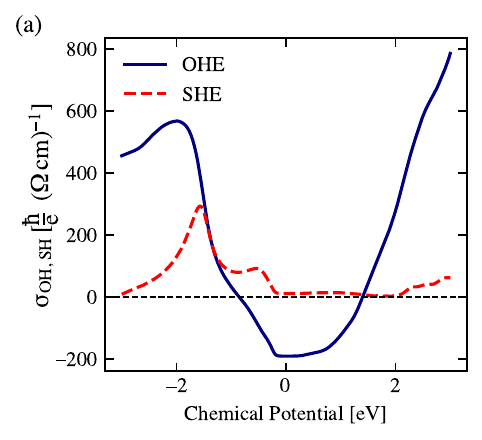}%
    \hspace{-2mm}%
    \includegraphics[width=0.38\textwidth]{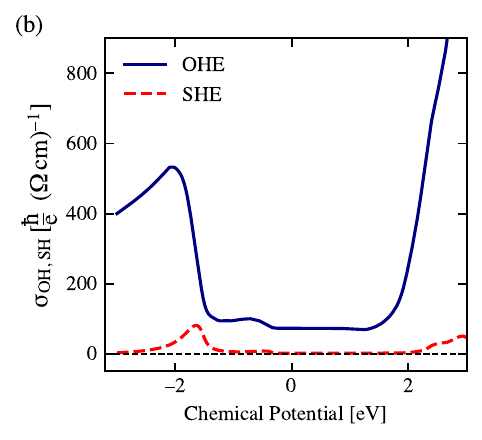}

    % SHE_OHE_GaAs_Si.ipynb
\caption{Intrinsic orbital Hall conductivity ($\sigma_\mathrm{OH}$, blue) and spin Hall conductivity ($\sigma_\mathrm{SH}$, red) of (a) GaAs and (b) Si as functions of the chemical potential.}
\label{fig:hall}
\end{widehere}

\subsection{Orbital and spin inverse Faraday effects}
\label{sec:ife_results}

Figure~\ref{fig:ife} shows the orbital and spin IFE coefficients of electron-doped GaAs and Si for the two light helicities. In both materials, the orbital coefficient is {much} larger in magnitude than the spin coefficient over most of the displayed photon-energy range. The GaAs response is also substantially larger in magnitude than the Si response. The divergence at vanishing photon energy is due to the prefactor $1/\omega^2$ in Eq.\ (\ref{eq:ife_terms}).

\begin{widehere}{figure}
    \centering    \includegraphics[width=0.37\textwidth]{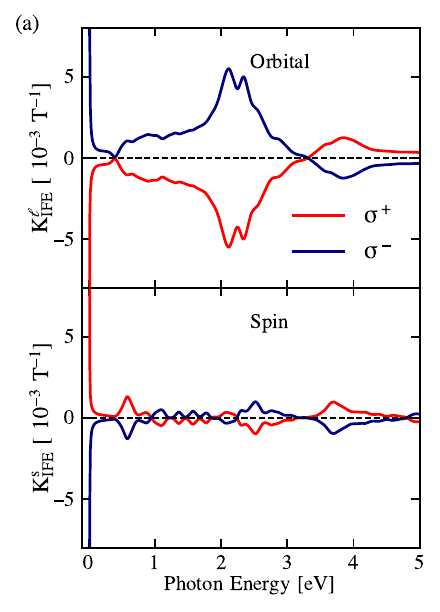}%
    \hspace{-2mm}   \includegraphics[width=0.37\textwidth]{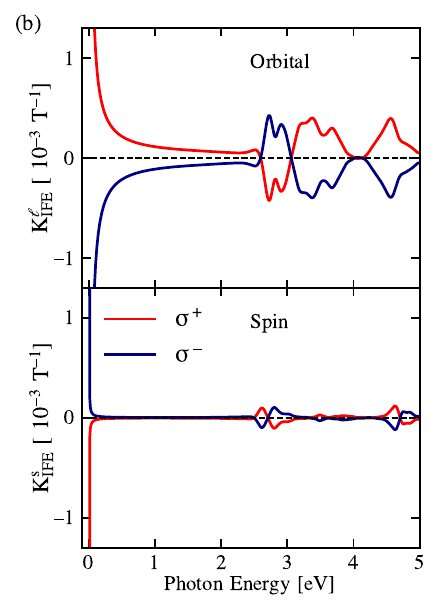}
\caption{Orbital (top) and spin (bottom) inverse Faraday coefficients $K_{\mathrm{IFE}}(\omega)$ of electron-doped (a) GaAs and (b) Si as functions of photon energy. Red and blue curves correspond to the two opposite light helicities $\sigma^\pm$.}
\label{fig:ife}
\end{widehere}

For a fixed helicity, the orbital and spin contributions have the same sign around the strongest GaAs feature and therefore add. In Si, their relative sign changes with photon energy. The two contributions can therefore either reinforce or partly cancel each other. Photon-energy tuning changes not only the magnitude of the total induced magnetization, but also its orbital and spin composition.

The much smaller spin response in Si than in GaAs is consistent with the weaker SOC of Si. At the same time, the sizable orbital response in both materials shows that the orbital and spin channels do not scale together. 
%\blue{This disparate behavior of the orbital and spin components emphasizes the non-Zeeman character of the optomagnetic field.}
Berritta \textit{et al.}\ showed that removing SOC in Fe suppresses the spin IFE while leaving a finite orbital contribution \cite{berritta2016ab}. Although that calculation concerned a different material, it demonstrates that the spin channel relies heavily on SOC \cite{freimuth2016laser} whereas the orbital channel does not.

A finite IFE in centrosymmetric Si is allowed because both the induced magnetization and $i\bm E\times\bm E^{*}$ are axial vectors and are unchanged by spatial inversion \cite{mishra2023spin}. For both nonmagnetic materials,
reversing the light helicity reverses the orbital and spin coefficients. The calculated $\sigma^{+}$ and $\sigma^{-}$ curves are opposite within numerical accuracy, consistent with the helicity-antisymmetric response of nonmagnetic systems~\cite{berritta2016ab}.

\begin{widehere}{table}
\caption{Dependence of the orbital, spin, and total IFE coefficients on
the broadening parameter $\Gamma_{nm}$. Results are evaluated at
$\hbar\omega=1.5$~eV for GaAs and $\hbar\omega=3.0$~eV for Si.
The coefficients are given in units of $10^{-3}$~T$^{-1}$. For each paired-sign entry ($\pm$ or $\mp$), the upper sign
corresponds to $\sigma^+$ and the lower sign to $\sigma^-$. The total coefficient is
$K_{\mathrm{IFE}}^{\mathrm{tot}}
=K_{\mathrm{IFE}}^{\ell}+K_{\mathrm{IFE}}^{s}$.}
\label{tab:ife_broadening}
\begin{ruledtabular}
\begin{tabular}{cccccc}

&
&

&
\multicolumn{3}{c}{
$K_{\mathrm{IFE}}\;(10^{-3}\,\mathrm{T}^{-1})$
}
\\[0.1cm]
\cline{4-6}\\[-0.3cm]
Material
&
$\hbar\omega$ (eV)
&
$\Gamma_{nm}$ (eV)
&
Orbital, $K_{\mathrm{IFE}}^{\ell}$
&
Spin, $K_{\mathrm{IFE}}^{s}$
&
Total, $K_{\mathrm{IFE}}^{\mathrm{tot}}$
\\[0.1cm]
\colrule\\[-0.3cm]
GaAs & 1.5 & 0.05 & $\mp 3.050$ & $\mp 0.390$ & $\mp 3.440$ \\
GaAs & 1.5 & 0.10 & $\mp 1.740$ & $\mp 0.318$ & $\mp 2.058$ \\
GaAs & 1.5 & 0.20 & $\mp 1.030$ & $\mp 0.203$ & $\mp 1.233$ \\[0.1cm]
\colrule\\[-0.3cm]
Si   & 3.0 & 0.05 & $\pm 0.100$ & $\mp 0.003$ & $\pm 0.097$ \\
Si   & 3.0 & 0.10 & $\mp 0.102$ & $\mp 0.013$ & $\mp 0.115$ \\
Si   & 3.0 & 0.20 & $\mp 0.152$ & $\mp 0.006$ & $\mp 0.158$ \\
\end{tabular}
\end{ruledtabular}
\end{widehere}

Table~\ref{tab:ife_broadening} presents the broadening energy dependence of the IFE coefficients at $\hbar\omega=1.5$~eV for GaAs and $\hbar\omega=3$~eV for Si. At these selected energies, the orbital contribution remains larger in magnitude than the spin contribution for all three broadening values, 0.05 eV, 0.1 eV, and 0.2 eV. In GaAs, reducing $\Gamma_{nm}$ increases the magnitudes of both contributions without changing their relative sign. For Si at $\hbar\omega = 3.0\,\mathrm{eV}$, the magnitude of the total IFE coefficient increases across the three broadening values considered. For each helicity, the total coefficient has the same sign at $\Gamma_{nm} = 0.10\,\mathrm{eV}$ and $0.20\,\mathrm{eV}$, and the opposite sign at $\Gamma_{nm} = 0.05\,\mathrm{eV}$. The orbital contribution remains dominant in all three cases.

\subsection{Moment-resolved optical Hall response and polar Kerr effect}
\label{sec:spin_orbital_Hall_response}

Figure~\ref{fig:source_sigma} shows the optical Hall coefficients per unit induced spin or orbital moment.

\begin{widehere}[0pt]{figure}
    \vspace*{12pt}
    \setlength{\abovecaptionskip}{2pt}
    \centering

    \includegraphics[
        width=0.37\textwidth,
        trim={0 5bp 0 5bp},
        clip
    ]{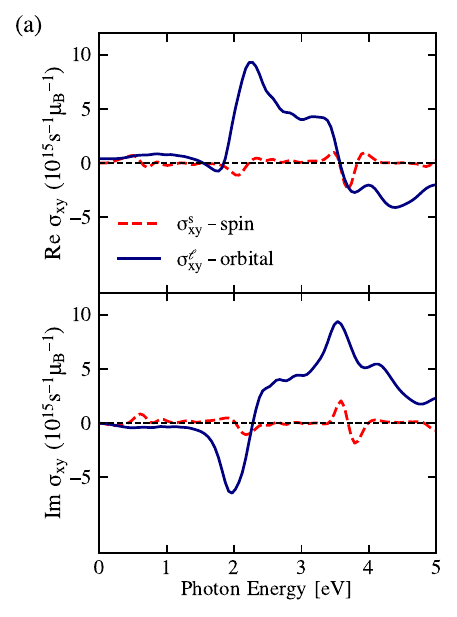}
    \hspace{-2mm}
    \includegraphics[
        width=0.37\textwidth,
        trim={0 5bp 0 5bp},
        clip
    ]{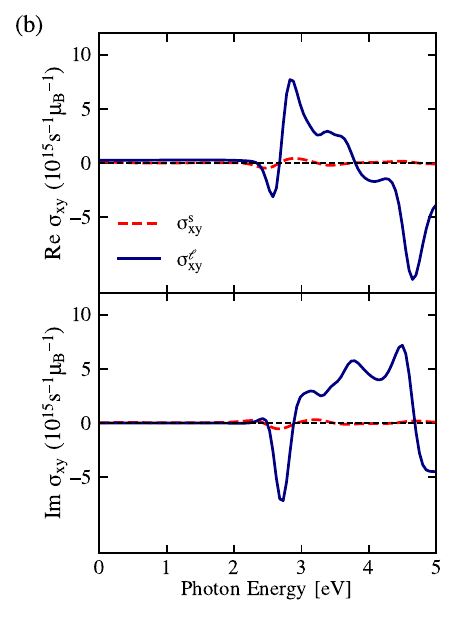}

    \caption{Real (top) and imaginary (bottom) parts of the moment-resolved optical Hall coefficients of electron-doped (a) GaAs and (b) Si as functions of photon energy. Blue solid and red dashed curves represent the orbital- and spin-source coefficients, denoted by $\sigma_{xy}^\ell$ and $\sigma_{xy}^s$, respectively.}
    \label{fig:source_sigma}
\end{widehere}

So far, we have investigated the spin and orbital responses generated by a dc electric field or circularly polarized light in the nonmagnetic ground states of GaAs and Si. We now examine the magneto-optical response produced by the resulting spin and orbital moments. 
The central result is that the optical Hall response is substantially more sensitive to an orbital moment than to a spin moment. The material dependence provides further insights. In Si, $\sigma_{xy}^{s}$ is strongly suppressed, whereas $\sigma_{xy}^{\ell}$ remains of the same order of magnitude as in GaAs. This trend is consistent with the established magneto-optical mechanism: electric-dipole transitions couple the helicity of light directly to the orbital degree of freedom, whereas a spin polarization becomes visible in the magneto-optical response through SOC~\cite{lyalin2023magneto,choi2023observation,argyres1955theory}. The stronger SOC in GaAs therefore helps explain its larger spin-source response, although the orbital contribution remains dominant over most of the investigated range.

The moment-normalized P-MOKE spectra $\Phi_K^\ell(\omega)$ and $\Phi_K^s(\omega)$ in Fig.~\ref{fig:bulk_kerr} retain the orbital-spin hierarchy found in the optical Hall response: over most of the investigated photon-energy range, an orbital moment produces a larger Kerr response than an equal spin moment. In the next section, we use these coefficients to calculate the spatial Kerr signal from the orbital and spin moments accumulated at the channel edges by the Hall currents.

\begin{widehere}{figure}
    \centering    \includegraphics[width=0.396\textwidth]{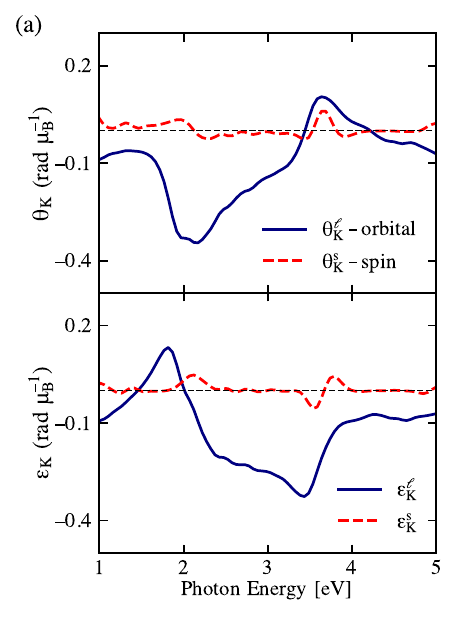}%
    \hspace{-2mm}%
    \includegraphics[width=0.396\textwidth]{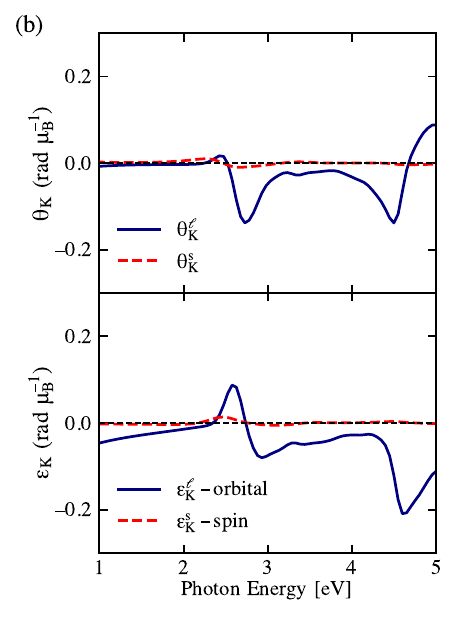}
    \caption{Real ($\theta_K^{\ell,s}$, top) and imaginary ($\varepsilon_K^{\ell,s}$, bottom) parts of the moment-resolved P-MOKE coefficients $\Phi_K^{\ell,s}$ of electron-doped (a) GaAs and (b) Si as functions of photon energy. Blue solid and red dashed curves represent the orbital- and spin-source coefficients, respectively.}
    \label{fig:bulk_kerr}
\end{widehere}

\subsection{Hall-induced lateral polar Kerr response}
\label{sec:Hall_induced_Kerr}

Finally, we investigate the expected lateral Kerr profiles generated by the SHE and OHE by combining the corresponding intrinsic Hall conductivities with a one-dimensional drift-diffusion model for the edge accumulations and the moment-normalized P-MOKE coefficients obtained above. We use $\hbar\omega=1.5$~eV for GaAs, corresponding to the $825$-nm probe used in Kerr-microscopy measurements on $n$-GaAs~\cite{kato2004observation}. For Si, we use $\hbar\omega=3.0$~eV, chosen above its calculated direct band gap of approximately $2.55$~eV. At these probe energies, the complex Kerr coefficients for GaAs are
\begin{align}
    \Phi_{K}^{\ell}(\hbar \omega = 1.5 \ \mathrm{eV})    &\simeq(-0.072+0.035i)~\mathrm{rad}\,\mu_{\mathrm B}^{-1},
    \label{eq:gaas_kerr_l}\\
    \Phi_{K}^{s}(\hbar \omega = 1.5 \ \mathrm{eV})
    &\simeq(0.0145-0.0034i)~\mathrm{rad}\,\mu_{\mathrm B}^{-1}.
\end{align}
For Si, the corresponding values are
\begin{align}
    \Phi_{K}^{\ell}(\hbar \omega = 3 \ \mathrm{eV})
    &\simeq(-0.041-0.072i)~\mathrm{rad}\,\mu_{\mathrm B}^{-1},
    \label{eq:si_kerr_l}\\
    \Phi_{K}^{s}(\hbar \omega = 3 \ \mathrm{eV})
    &\simeq(-0.0032-0.005i)
    ~\mathrm{rad}\,\mu_{\mathrm B}^{-1}.
\end{align}

The real and imaginary parts are the Kerr rotation and ellipticity coefficients, respectively.
We assume a thin film geometry of width $W=80~\mu\mathrm{m}$ along $x$, with an electric field $E_y$ applied along $y$.  The Hall currents flow along $x$ and carry $z$-polarized orbital or spin angular momentum, producing opposite moment accumulations at the two lateral edges. Following the drift-diffusion treatment used for the OHE-induced accumulation in Ti~\cite{choi2023observation}, the corresponding accumulated-moment profile is given by
\begin{equation}
    m_q(x)=m_{0,q}
    \frac{\sinh(x/l_q)}
    {\cosh[W/(2l_q)]},
    \qquad q=\ell,s,
    \label{eq:lateral_moment_profile}
\end{equation}
where $m_q(x)$ is the $z$-directed accumulated orbital or spin magnetic moment per two-atom primitive cell, $m_{0,q}$ is its amplitude, and $l_q$ is the corresponding relaxation length.

The condition of zero transverse angular-momentum current at
$x=\pm W/2$ gives
\begin{equation}
	m_{0,q}=\frac{a^3}{4}\gamma_q\sigma_\mathrm{OH/SH} l_q E_y / D_q,
    \label{eq:moment_amplitude}
\end{equation}
where $\sigma_\mathrm{OH/SH}$ is the orbital or spin Hall conductivity, $\gamma_\ell$ and $\gamma_s$ denote the orbital and spin gyromagnetic ratios, and $D_q$ is the diffusion constant for $q = \ell, s$.
%$\gamma_\ell=-\mu_{\mathrm B}/\hbar$ and $\gamma_s=-2\mu_{\mathrm B}/\hbar$ are the signed orbital and spin gyromagnetic ratios, respectively.
For the illustrative profiles, we adopt the baseline modeling approximation
\begin{equation}
    D_s\simeq D_\ell\simeq D_c .
    \label{eq:diffusion_approximation}
\end{equation}
This assumption is not intended to imply that the microscopic spin, orbital, and charge diffusion constants are generally identical; instead, we aim to roughly predict the reasonable P-MOKE signals that can be measured in experiments.

To obtain experimentally relevant estimates, we adopt representative room-temperature electron diffusion coefficients based on the doping-dependent results of Ref.~\cite{scali2025frequency} for $n$-doped Si and GaAs at carrier densities of order $10^{18}~\mathrm{cm^{-3}}$:
\begin{equation}
D_q^{\mathrm{Si}}=7~\mathrm{cm^2 s^{-1}},
\qquad
D_q^{\mathrm{GaAs}}=70~\mathrm{cm^2 s^{-1}}.
\label{eq}
\end{equation}

The resulting complex local Kerr response is given by $\Phi_K^{q}m_q(x)$, where $\Phi_K^{q}$ is the complex P-MOKE coefficient per unit orbital or spin magnetic moment, evaluated at the probe energy specified above for each material. The corresponding Kerr rotation is then obtained as
\begin{equation}
\theta_K^{q}(x)
    =\operatorname{Re}\!\left[
\Phi_K^{q}m_q(x)\right].
\label{eq:kerr_profile}
\end{equation}

% At $T=300~\mathrm{K}$ and for the carrier concentrations used here,
% \begin{equation}
%     n_{\mathrm{Si}}
%     =9.99\times10^{17}~\mathrm{cm^{-3}},
%     \qquad
%     n_{\mathrm{GaAs}}
%     =8.86\times10^{17}~\mathrm{cm^{-3}},
% \end{equation}

Figure \ref{fig:kerr_model} shows the estimated position-dependent Kerr rotation signals originating from the OHE and SHE, where we use $E_y=10~\mathrm{V\,cm^{-1}}$ and the characteristic lengths are chosen as $l_q=5$, $3$, $1~$, and $0.5~\mu\mathrm{m}$. Using the same lengths allows the profile shapes to be compared directly; it does not mean that the physical orbital and spin relaxation lengths are equal.

\begin{widehere}{figure}
    \centering
    \includegraphics[width=0.7\textwidth]{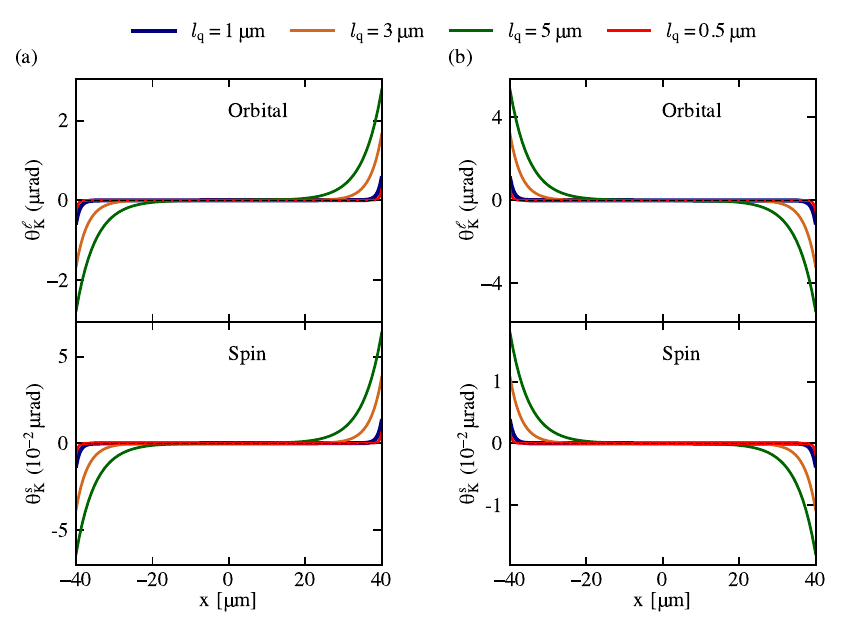}

\caption{Calculated Hall-induced lateral P-MOKE rotation profiles at $\hbar\omega=1.5~\mathrm{eV}$ for (a) GaAs and $\hbar\omega=3~\mathrm{eV}$ for (b) Si. The upper and lower panels show the orbital- and spin-Hall-induced Kerr rotations, denoted by $\theta_K^\ell$ and $\theta_K^s$, respectively. We use $W=80~\mu\mathrm{m}$ and $E_y=10~\mathrm{V\,cm^{-1}}$. The curves correspond to different assumed diffusion lengths, $l_q=0.5$, $1$, $3$, and $5~\mu\mathrm{m}$, where $q=\ell,s$. Note that the vertical scale of the spin panels is two orders of magnitude smaller than that of the orbital panels.}   
\label{fig:kerr_model}
\end{widehere}

We note that although the orbital relaxation lengths in transition metals are typically limited to the nanometer scale \cite{lee2021efficient, rang2024orbital,guan2026evidences,hayashi2023observation,guan2026evidences}, those for the lateral orbital transport in weakly doped semiconductors have not been reported, to the best of our knowledge. The results show antisymmetric, edge-localized Kerr profiles, reflecting the opposite Hall accumulations at the two channel boundaries.
Within the adopted approximations, the orbital contribution is much larger than the spin contribution in both materials because both the calculated orbital Hall conductivity and the Kerr response per unit orbital moment exceed their spin counterparts.  Consequently, our results indicate that the current-induced P-MOKE signals are expected to be dominated by the OHE over the SHE, within the intrinsic regime.

\section{Conclusion}

Our first-principles calculations for electron-doped GaAs and Si yield intrinsic orbital Hall conductivities that exceed their spin counterparts in magnitude. The IFE also produces predominantly orbital magnetization over broad photon-energy intervals. In both GaAs and Si, reversing the light helicity reverses both the spin and orbital contributions to the IFE magnetization. These results establish substantial orbital contributions to both electrically driven angular-momentum transport and optically induced magnetization.

Optical detection introduces a further distinction between spin and orbital moments. For moments of equal magnitude, the calculated optical Hall and P-MOKE responses are larger in magnitude for the orbital contribution over most of the investigated photon-energy range. At the selected photon energies, the magnitude of the Kerr rotation per unit orbital moment is approximately 5 times its spin counterpart in GaAs and 13 times in Si. The sizable orbital optical response in Si, despite its weak spin response, is consistent with direct electric-dipole coupling to orbital motion and SOC-mediated optical sensitivity to spin polarization.

The drift-diffusion calculation connects Hall transport with optical detection by combining the intrinsic Hall conductivities with the moment-resolved P-MOKE coefficients. It gives antisymmetric Kerr profiles with predominantly orbital contributions at the channel edges under the assumptions of equal diffusion coefficients and relaxation lengths for spin and orbital moments. The predominance of the orbital Kerr contribution, {being nearly two orders of magnitude larger,} reflects both the larger intrinsic orbital Hall conductivity and the stronger optical response per orbital moment. The predicted amplitudes remain conditional on these transport assumptions. These findings provide a basis for distinguishing spin and orbital contributions to current-induced magneto-optical signals and motivate further experimental investigations of orbital transport in semiconductors.
\\

\section*{Acknowledgments} We thank Marco Berritta for helpful discussions.
This work has been supported by the Swedish Research Council (VR) and the Wallenberg Initiative Materials Science for Sustainability (WISE) funded by the Knut and Alice Wallenberg Foundation. We further acknowledge funding from the Knut and Alice Wallenberg Foundation (Grant No.\ 2022.0079 and No.\ 2023.0336) and the EIC Pathfinder OPEN “OBELIX” (Grant No.\ 101129641). The calculations were supported by resources provided by the National Academic Infrastructure for Supercomputing in Sweden (NAISS) at NSC Linköping, partially funded by Vetenskapsrådet through Grant No.\ 2022-06725 and No.\ 2026-05211. 
% \clearpage
% \bibliographystyle{apsrev4-1}
% \IfFileExists{ref.bib}{\bibliography{ref}}{\bibliography{../ref}}

%\bibliographystyle{apsrev4-2}
\bibliography{ref.bib}

%\newpage
\end{document}